%% file: draftfvflores.tex
\documentclass[a4paper,10pt,onecolumn]{article}

\newcommand{\sh}[1]{#1\hskip-7pt \diagup}

\usepackage{here}

\usepackage{latexsym}
\usepackage{amssymb}
\usepackage{mathrsfs}
\usepackage{amsfonts}

\usepackage[utf8x]{inputenc}
\usepackage{graphicx}
\usepackage{epsfig}
\usepackage{amsmath}
\title{Model Independent  Electromagnetic corrections in hadronic $\tau$ decays}
\author{F.V. Flores-Ba\'ez,  J.R. Morones-Ibarra.\\ \small{Facultad de Ciencias F\'isico-Matem\'aticas, Universidad Aut\'onoma de Nuevo Le\'on}
\\ \small{Ciudad Universitaria, San Nicol\'as de los Garza, Nuevo Le\'on, 66450, M\'exico}}

\begin{document}

\maketitle

\begin{abstract}
The  long distance correction to the total  decay width of the $\tau^{\pm}\to K^{0}\pi^{\pm}\nu$ decay is calculated in a model independent approach where a
discrimination of photons in the Bremmstrahlung process is assumed. This correction is completely free of UV and IR singularities and moreover, it satisfies
electromagnetic gauge invariance. The result of this work can be applied on the tau decays: $\tau^{\pm}\to \pi^{\pm}\pi^{0}\nu, K^{\pm}\pi^{0}\nu$.  
\end{abstract}

\section{Introduction}
It is well known that hadronic $\tau$ decays are an ideal laboratory to obtain information about the fundamental parameters within the Standard Model and also 
some properties of QCD at low energies \cite{sm}\cite{qcd}. In particular the $\tau\to K\pi \nu$ decay has been studied in the past by ALEPH \cite{aleph} and 
OPAL \cite{opal} and recently at B factories \cite{babar}\cite{belle} where the high statistic measurements provide excellent information about the structure of 
the spectral functions, parameters of the intermediate states and the total hadronic spectral function. It is also possible to determine the product 
$|V_{us}|F_{+}^{K^{0}\pi^{+}}(0)$ from this decay although the best determination comes from semileptonic kaon decays \cite{antonelli}. 

On the theoretical side this sort of processes have a nice feature: the decay amplitude can be factorized into a pure leptonic part and a hadronic spectral 
function \cite{mirkes} in such a way that the differential decay distribution reads
\begin{align}\label{eq:pich}
\frac{d\Gamma_{K\pi}}{d\sqrt{s}}&=\frac{G_{F}^{2}\left(|V_{us}|F_{+}^{K\pi}(0)\right)^{2}m^{3}_{\tau}}{32\pi^{3}s}S_{EW}\left(1-\frac{s}{m^{2}_{\tau}}\right)^{2}
\nonumber\\
&\times\left[\left(1+\frac{2s}{m^{2}_{\tau}}\right)q^{3}_{K\pi}|\tilde{F}_{+}^{K\pi}(s)|^{2}
+\frac{3\Delta_{K\pi}^{2}q_{K\pi}|\tilde{F}_{0}^{K\pi}(s)|^{2}}{4s}\right]\ ,\nonumber\\
\end{align}
where a sum over the two posibles decays $\tau^{+}\to K^{+}\pi^{0}\nu$ and $\tau^{+}\to K^{0}\pi^{+}\nu$ has been done, isospin symmetry is assumed and the reduced 
vector and scalar form factors have been normalized to one at the origin
\begin{align}
\tilde{F}_{+}^{K\pi}(s)&= \frac{F_{+}^{K\pi}(s)}{F_{+}^{K\pi}(0)}\ ,\qquad \tilde{F}_{0}^{K\pi}(s)= \frac{F_{0}^{K\pi}(s)}{F_{+}^{K\pi}(0)}\ .
\end{align}
In this expresion $\Delta_{K\pi}\equiv m^{2}_{K}-m^{2}_{\pi}$  and the kaon momentum in the rest frame of the hadronic system reads
\begin{align}
q_{K\pi}&=\frac{1}{2\sqrt{s}}\sqrt{(s-(m_{K}+m_{\pi})^{2})(s-(m_{K}-m_{\pi})^{2})}\times \theta(s-(m_{K}+m_{\pi})^{2})\ .
\end{align}
A theoretical description\cite{jamin} of the vector and scalar form factors has been done in the Chiral Theory with Resonances(R$\chi$T) framework\cite{rchpt},
providing a succesful representation of the data. It is worth to mention that  eq.(1) includes the short distance correction $S_{EW}$ \cite{sirlinmarciano}\cite{erler}
however the long distance correction is omitted. In the B factories\cite{babar}\cite{belle}  an improved experimental precision can be achieved in the future which makes
mandatory to have a  theoretical analysis of the long distance corrections effects.\\
It is well known that in all decays  with a charged particle the emission of photons is always present altering  the dynamics of the decay.
The approximative next-to-leading  order algorithms \cite{photos} are used to simulate the correction due to  soft photons where  the virtual  corrections (one loop) 
are  reconstructed numerically up to the leading logarithms from the real photon corrections. An improved algorithm \cite{klor} can be applied if a phenomenological
model is used to describe the behaviour of the invariant amplitude.\\
On the other hand, the first attempt to describe the lepton-hadron interaction used a simple effective interaction approach, nonetheless once the electromagnetic
correction is computed an unfortunate feature appears: it depends on the cutoff energy \cite{ginsbergI} that controls the UV singularity. Nevertheless in the $\chi$PT 
framework \cite{chpt} it is possible to describe the interactions between the lightest multiplet of pseudoscalar mesons and the lightest leptons  at low energy 
(below the $\rho(770)$ resonance region) including also real and virtual photons \cite{chptconfotones}\cite{chptleptons} and 
due to the character of $\chi$PT, the UV singularity is cancelled by adding a finite number of appropiated counterterms. At the end one does not have to deal
with an UV cutoff  but with the finite pieces of the couplings constant of the effective theory. A nice example of this type of electromagnetic correction treatment 
was done in  $K_{l3}$ \cite{kl3}.\\
Another alternative for the analysis of electromagnetic corrections is given by a Model Independent (MI) approach \cite{Yena}. In this method the invariant amplitude
of the radiative correction is separated into external and internal contribution. The external radiative correction is obtained by considering the emission and 
absorption of real or virtual photons in external lines.
The internal radiative correction corresponds to diagrams where a photon is attached to an internal line and clearly depends
on the precise details of the process, in other words this part is model dependent. The aim of this technique is to procure an electromagnetic correction, evaluated 
once and for all  which  is gauge invariant, free of UV singularities and contains the dominant logarithms that come from the infrared singularity cancellation.\\
The MI technique has been applied in the analysis of the electromagnetic corrections to the Dalitz Plot of semileptonic decays of hyperons \cite{augusto1}, in
the calculation of radiative correction to leptonic decays of pseudoscalar meson\cite{augusto2}  and also recentely a radiative corrections analysis to the Dalitz
plot of the $K_{l3}$ decay\cite{alfonso}. It is worth to mention that authors in \cite{alfonso}  have quoted that their result in  $K_{l3}$ has been compared with the 
universal electromagnetic correction given in \cite{kl3} with a very small quantitative difference. That comparison also help us to identify  the model dependent part in the MI 
scheme with the corresponding  form factor computed within $\chi$PT. In the  effective approach the form factors can be separated into two parts, one of them contains
the universal electromagnetic correction that is linked with the model independent electromagnetic factor (MI)  and the second one can be seen as the 
corresponding model dependent form factor (MD), which in the context of $\chi$PT is made of hadronic loop contributions, finite local terms and electromagnetic pieces 
that are left aside once the universal correction is defined\footnote{It is worth to mention that the definition of the universal e.m. correction is not unique meanwhile the e.m. correction
computed in the MI approach is very well defined.}.\\
In the hadronic $\tau$ decays the energy transfered $\sqrt{s}$ to the hadronic system goes from a treshold $(m_{1}+m_{2})$ up to $m_{\tau}$, which means that
the expansion parameter of $\chi$PT is no longer valid in all the region. In this respect we can not apply consistently the work done in $k_{l3}$ in computing the
electromagnetic corrections in hadronic tau decays.

In this paper we are not comitted in to give a description of the form factors, our aim is to present the MI electromagnetic corrections to the $\tau\to K\pi\nu$ 
following the techniques given in \cite{Yena}. In Section 2 we describe  briefly  the basis of the hadronic tau decay and  general effects of electromagnetic 
correction. In Section 3 the MI electromagnetic corrections for virtual and real photons are computed. In Section 4 we present the MI corrections to the
 $\tau^{+}\to K^{0}\pi^{+}\nu$ decay width and a discussion about our result.

\section{$\tau$ decay and the Electromagnetic Interaction}

\subsection{Basic elements on $\tau$ decays}
Here we do a brief review of  the very well known elements of the $\tau^{\pm}\to  P^{\pm} P^{0}\nu_{\tau}$ decay where $P$ is a pseudoscalar meson. We start with the 
invariant amplitude without electromagnetic effects that is expressed \cite{mirkes} as follows
\begin{eqnarray}\label{eq:mglevel}
\mathcal{M}^{(0)}_{\tau}&=& C_{CG}G_{F}V_{CKM}l_{\mu}h^{\mu}\ ,
\end{eqnarray}
where $G_{F}$ is the Fermi constant, $V_{CKM}$ is the CKM matrix element, the Clebsch-Gordon
coefficient $C_{CG}=(-\frac{\sqrt{2}}{2},\frac{1}{2},1)$ depends on the hadronic final state ($\pi^{+}K^{0}, \pi^{0}K^{+},\pi^{+}\pi^{0}$)
and the leptonic current is given by\footnote{We use the notation $\gamma^{7}= 1-\gamma^{5}$.} 
\begin{eqnarray}
l_{\mu}&=&\bar{u}_{\nu_{\tau}}(q)\gamma_{\mu}\gamma^{7}u_{\tau}(p_{\tau})\ ,
\end{eqnarray}
meanwhile the hadronic part $h^{\mu}$  that contains the form factors $F_{+}(s)$, $ F_{-}(s)$ is  written as follows
\begin{eqnarray}\label{eq:htff}
h^{\mu}=F_{+}(s)(p^{+}-p^{0})^{\mu}+F_{-}(s)(p^{+}+p^{0})^{\mu}\ ,
\end{eqnarray}
where $ p_{+}(p_{0})$ is the 4-momenta of the charged (neutral) pseudo-scalar meson and $s=(p_{+}+p_{0})^{2}$.\\
The density function at tree level is given by
\begin{align}\label{eq:density}
\rho^{(0)}(s,u)&= |F_{+}(s)|^2\mathcal{D}(s,u)+2F_{+}(s)F^{\dagger}_{-}(s)\mathcal{D}_{3}(s,u) +|F_{-}(s)|^2\mathcal{D}_{2}(s) \ ,
\end{align}
and the $\mathcal{D}$-functions are given as follows 
\begin{align}
D(s,u)&= 2u^2+2u(s-m^2_{\tau}-m^2_{+}-m^2_{0})+\frac{m^2_{\tau}}{2}(m^2_{\tau}-s)+2m^2_{+}m^2_{0} \ ,\nonumber\\
D_{2}(s)&=\frac{m^2_{\tau}}{2}(m^2_{\tau}-s)\ ,\nonumber\\
D_{3}(s,u)&=\frac{m^2_{\tau}}{2}(2m^2_{0}+m^2_{\tau}-s-2u)\ .
\end{align}
The differential decay width of the  process $\tau^{+}\to P^{+}P^{0}\nu_{\tau}$ can be written in a simple form
\begin{align}
\frac{d\Gamma^{(0)}_{P^{+}P^{0}}}{dsdu}&=\frac{|C_{CG}G_{F}V_{CKM}|^2}{(4\pi)^3m^3_{\tau}}\rho^{(0)}(s,u)
\end{align}
where $u=(p_{\tau}-p^{+})^2$. After doing the u-integration 
it is straightforward to get\footnote{$\Delta_{+}=m^2_{+}-m^2_{0}$ where $m_{+}(m^2_{0})$ is the mass of the charged(neutral)  meson.}
\begin{align}\label{eq:dtd}
\frac{d\Gamma^{(0)}_{P^{+}P^{0}}}{ds}&=\frac{|C_{CG}G_{F}V_{CKM}|^2m^3_{\tau}q_{+0}(1-\frac{s}{m^2_{\tau}})^2}{3(4\pi)^3\sqrt{s}}\left\{ 3\left|F_{-}(s)\right|^2
\right.\nonumber\\&\left. +|F_{+}(s)|^2\left[(1+\frac{2s}{m^2_{\tau}})\frac{4q^{2}_{+0}}{s}+\frac{3\Delta_{+}^2}{s^2}\right]+6 F_{+}(s)F^{\dagger}_{-}(s)\frac{\Delta_{+}}{s}\right\} \ ,
\end{align}
where the  momentum $q_{+0}$ in the rest frame of the hadronic system reads
\begin{align}
q_{+0}&= \frac{1}{2\sqrt{s}}\sqrt{(s-(m_{+}+m_{0})^2)(s-(m_{+}-m_{0})^2)}\times \theta(s-(m_{+}+m_{0})^2)\ .
\end{align}
It is more familiar and elegant to write eq.(\ref{eq:dtd}) in terms of the scalar form factor $F_{0}(s)$ defined by  
\begin{align}\label{eq:dsff}
F_{-}(s)=\frac{\Delta_{+}}{s}[F_{0}(s)-F_{+}(s) ]\ .
\end{align}
With the help of eq.(\ref{eq:dsff}), the differential decay width reads
\begin{align}
 \frac{d\Gamma^{(0)}_{P^{+}P^{0}}}{d\sqrt{s}}&=\frac{|G_{F}V_{CKM}|^2m^3_{\tau}}{32\pi^3 s}[1-\frac{s}{m^2_{\tau}}]^2\left[ \frac{3\Delta_{+}^2q_{+0}}{4s}|F_{0}(s)|^2
\right.\nonumber\\&\left. +|F_{+}(s)|^2 (1+\frac{2s}{m^2_{\tau}})q_{+0}^3\right]A_{p^{+}p^{(0)}}\ ,
\end{align}
where $A_{p^{+}p^{(0)}}=\frac{4}{3}C^2_{CG}$ and its value is equal to  $\frac{1}{3}(\frac{2}{3})$  for the hadronic final state $K^{+}\pi^{0},(K^{0}\pi^{+})$. It is 
easy to see that summing both final $K\pi$ states we get eq.(\ref{eq:pich}). In the case of the $\pi^{+}\pi^{0}$-final state, $A_{\pi\pi}=4/3$ and the scalar form 
factor contribution vanishes\footnote {Even considering the physical pion masses, the scalar contribution is negligible.} due to isospin symmetry.\\
The energy-momenta conservation defines a set of allowed values for the cinematic variables $(u,s)$ that can be shown in a Dalitz plot which borders  are given by
\begin{align}\label{fase1}
(m_{+}+m_{0})^{2}\leq s \leq m^{2}_{\tau}\ ,\qquad u_{-}(s)\leq u \leq u_{+}(s)\ ,
\end{align}
where
\begin{align}\label{fase2}
u_{+}(s)&=\frac{1}{2s}\left\{  2s(m^{2}_{\tau}+m^{2}_{0}-s)-(m^{2}_{\tau}-s)(s+m^{2}_{+}-m^{2}_{0})\right.\nonumber\\
&\left. +(m^{2}_{\tau}-s)\sqrt{(s-(m_{+}^{2}+m_{0}^{2}))^{2}-4m^{2}_{+}m^{2}_{0}}  \right\}\ ,\nonumber\\
u_{-}(s)&=\frac{1}{2s}\left\{  2s(m^{2}_{\tau}+m^{2}_{0}-s)-(m^{2}_{\tau}-s)(s+m^{2}_{+}-m^{2}_{0})\right.\nonumber\\
&\left. -(m^{2}_{\tau}-s)\sqrt{(s-(m_{+}^{2}+m_{0}^{2}))^{2}-4m^{2}_{+}m^{2}_{0}}\right\}\ .
\end{align}
Considering the hadronic final state $K^{0}\pi^{+}$, the integrated decay width is found to be
\begin{align}
\Gamma^{(0)}_{K\pi}&=|F_{+}(0)V_{CKM}|^2G_{F}^2 C^2_{CG}\mathcal{N}_{i}I^{(0)}_{K\pi}
\end{align}
where $\mathcal{N}_{i}=\left[\frac{m^5_{\tau}}{48\pi^3}\right]$ and the integrated density $I^{(0)}_{K\pi}$ is defined as follows
\begin{align}
I_{K\pi}^{(0)}&= \int \frac{ds}{m^2_{\tau}s\sqrt{s}}[1-\frac{s}{m^2_{\tau}}]^2\left[ |\tilde{F}^{K\pi}_{+}(s)|^2 (1+\frac{2s}{m^2_{\tau}})q_{K\pi}^3 \right.\nonumber\\&\left.
+\frac{3\Delta_{+}^2 q_{K\pi}}{4s}|\tilde{F}^{K\pi}_{0}(s)|^2 \right]\ .
\end{align}
In order to compute $I_{K\pi}^{(0)}$ one needs a theoretical description of the normalized vector and scalar form factors $\tilde{F}^{K\pi}_{+}(s)$ and 
$\tilde{F}^{K\pi}_{0}(s)$. This task is achieved by a fit to the measured distributions to  $\tau^{\pm}\to K_{s}\pi^{\pm}\nu$. There is a strong effort in computing 
this integral \cite{emiliework} with a dispersive representation of the form factors but for illustrative purpose  we consider the parameterized vector and scalar 
form factors as given by the Belle Collaboration \cite{belle} as follows
\begin{align}
F_{V}(s)&=F_{+}(s)=\frac{1}{1+\beta}[BW_{K^{*}(892)}(s)+BW_{K^{*}(1410)}(s)]\ ,
\end{align}
where $\beta$ is the fraction of the $K^{*}(1410)$ resonance contribution and $BW(s)_{R}$ is a relativistic Breit-Wigner function
\begin{align}
BW_{R}(s)&=\frac{M^2_{R}}{M^2_{R}-s-\imath M_{R}\Gamma_{R}(s)}\ ,
\end{align}
and $\Gamma_{R}(s)$ is the s-dependent total width of the resonance,
\begin{align}
\Gamma_{R}(s)&= \Gamma_{0,R}\frac{s}{M^2_{R}}\left(\frac{\sigma_{K\pi}(s)}{\sigma_{K\pi}(m^2_{R})}\right)^3\ ,\nonumber\\
\sigma_{K\pi}(s)&=\frac{2q_{K\pi}(s)}{\sqrt{s}}\ .
\end{align}
For the scalar form factor $F_{0}(s)$ we take the description that includes only the $K^{*}_{0}(800)$ resonance
\begin{align}
F_{0}(s)&=\kappa \frac{s}{M^{2}_{K^{*}_{0}(800)}}BW_{K^{*}_{0}(800)}(s)\ .
\end{align}
where $\kappa$ is a complex constant and  represents the fraction of the scalar resonance contribution.\\
The mass and width of $K^{*}(892),K^{*}(1410)$ are fixed from \cite{pdg} meanwhile the parameters of $K_{0}^{*}(800)$ are taken from \cite{ablikim}. With 
the values of Table 3  given in\cite{belle}, it  is found  that $I_{K_{S}\pi^{\pm}}^{(0)}= 0.384221$. Assuming that
\begin{align}
\mathcal{B}(\tau^{\pm}\to K^{0}\pi^{\pm}\nu)&=\mathcal{B}(\tau^{\pm}\to K_{S}\pi^{\pm}\nu)+\mathcal{B}(\tau^{\pm}\to K_{L}\pi^{\pm}\nu)\nonumber\\
&= 2\mathcal{B}(\tau^{\pm}\to K_{S}\pi^{\pm}\nu)\ ,
\end{align}
then we get 
\begin{align}
I^{(0)}_{K^{0}\pi{\pm}}=2I^{(0)}_{K_{S}\pi{\pm}}\sim 0.768\ .
\end{align}

\begin{figure}[hbt]
\centerline{\includegraphics[width=.95\linewidth]{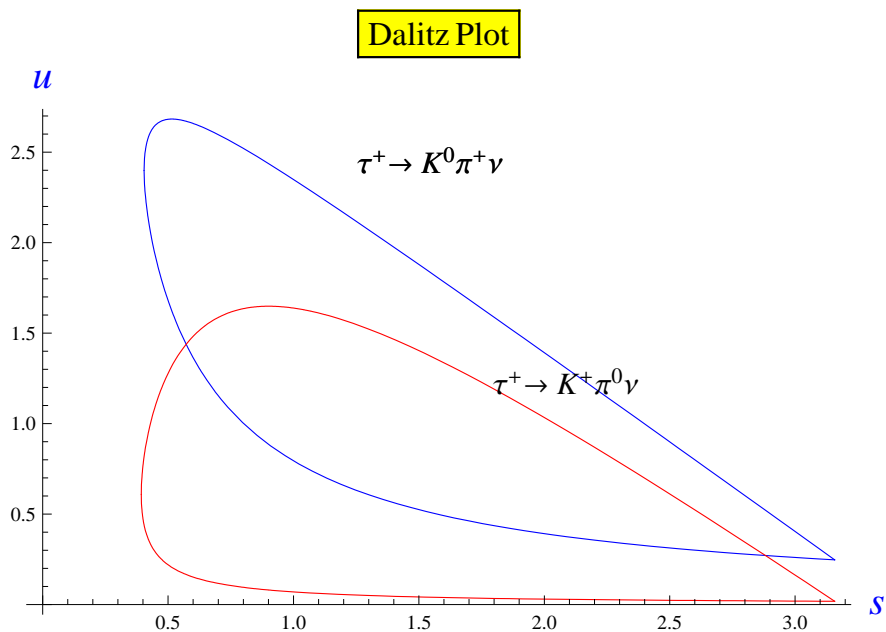}}
\caption{\footnotesize The Dalitz plot without electromagnetic corrections for the tau decay into the $K\pi$ hadronic final state, the Mandelstand variables
are defined in the text and masses are  given in GeV.}
\label{fig1}
\end{figure}

\subsection{ An overview on electromagnetic corrections}
Here we addres some general effects of the electromagnetic corrections. First we assume the simplest and easiest scenario: there is just one  form factor that  
is not affected by the one loop integration, in other words its dependence on s is negligible and we denote this by writing $F_{+}(s)=F_{V}$. In this case  the tree 
level amplitude for the $\tau^{+}\to P^{+}P^{0}\nu  $ decay reads
\begin{eqnarray}\label{eq:tlevel}
\mathcal{M}^{(0)}=G_{F}V_{CKM}C_{CG}F_{V}l_{\mu}(p_{+}-p_{0})^{\mu}\ .
\end{eqnarray}
The  amplitude for the one loop electromagnetic correction with pointlike meson-photon interaction is found to be
\begin{equation}
\mathcal{M}^{v}=G_{F}V_{CKM}C_{CG}F_{V} l_{\mu}\frac{\alpha}{4\pi}\left[f^{e.m.}_{+}(u)t^{\mu}_{-}+f^{e.m.}_{-}(u)t^{\mu}_{+} \right]\ ,
\end{equation}
where $t^{\mu}_{\pm}=(p^{+} \pm p^{0})^{\mu}$. It is well known that the function $f^{e.m.}_{+}(u)$ encloses ultraviolet (UV) and infrared (IR) singularities meanwhile 
$f^{e.m.}_{-}(u)$ contains  just  an UV singularity. The IR divergence  is cancelled by taking into account the real photon emission meanwhile the UV problem can be 
solved with a cut-off in the one-loop integration that replaces the singularity\footnote{$D= 4-\epsilon$ assuming  the one loop integrals are
done in the Dimensional Regularization Method, however anothers prescriptions are equivalents.}. In short this method consists in making the next replacement  
\begin{align}
\frac{2}{4-D}-\gamma_{E}+\ln[4\pi]+ \ln [\frac{m^2}{\mu^2}]\rightarrow \ln[\frac{\Lambda^2}{m^2}]\ .
\end{align}
where  the cut-off $\Lambda$ can be some resonance mass and $m$ is a caracteristic mass of the process. The replacement makes sense because 
represents that something is hidden in the effective 4-body interaction encoded in eq.(\ref{eq:tlevel}).\\
In order to make an analogy we recall  the one loop corrections (at short distances) computed \cite{braaten} within the Standard Model for the process $\tau\to \nu_{\tau} d\bar{u}$ where  
the pure QED correction in the Fermi model is UV divergent but if all the electroweak corrections are included ( $Z$ and $W$ virtual exchange) then the ultraviolet 
cut-off ( ln[$\frac{\Lambda^2}{m^2_{\tau}}$]) is replaced naturally  by a large logarithm namely $\ln [m_{Z}/m_{\tau}]$\cite{sirlinmarciano}. In this case the hidden
objects in the Fermi model  are $Z$ and  $W$ bosons.\\
On the other hand, when the electromagnetic corrections are computed in the MI approach, the  discussion about the value of the cut-off $\Lambda$ is left aside.
For instance consider we compute the e.m. corrections to eq.(\ref{eq:tlevel})  following the work of \cite{Yena},
in this case the one loop invariant matrix can be written as follows
\begin{align}\label{eq:ol}
\mathcal{M}^{v}&= \mathcal{M}^{(0)}\frac{\alpha}{4\pi} f^{e.m.}_{m.i.}(u,\lambda^2)+ \mathcal{M}_{+} \frac{\alpha}{4\pi}f^{r}_{m.i.}(u)
+\mathcal{M}^{(0)}\frac{\alpha}{4\pi}f^{m.d}_{+}(u)\ ,
\end{align}
where 
\begin{align}\label{eq:exmia}
\mathcal{M}_{+}&=G_{F}V_{CKM}C_{CG}F_{V}\bar{u}_{\nu_{\tau}}(q)\gamma_{\mu}\gamma^{7}\sh{p}^{+} u_{\tau}(p_{\tau})t^{\mu}_{-}\ .
\end{align}
The first and second pieces of eq.(\ref{eq:exmia}) are the e.m. corrections independent of structure effects, come from the convection and
spin term, are  gauge invariant, free of UV divergences and
all the IR singularity is located in this term. The last piece  gathers in the function $f^{m.d}_{+}(u)$, the e.m. effects on the vector form factor, details of strong interactions 
and the model dependent assumptions to cancel the UV divergences. Notice that in assuming that the vector form factor is the dominant we are taking into 
account only the effects at $\mathcal{O}(\alpha)$ to this form factor. Adding  eqs.(\ref{eq:tlevel},\ref{eq:ol}) we get
\begin{align}\label{eq:ctl}
\mathcal{M}&=\mathcal{M}^{(0)'} [1+\frac{\alpha}{4\pi} f^{e.m.}_{m.i.}(u,\lambda^2)]+ \mathcal{M}_{+}^{'} \frac{\alpha}{4\pi}f^{r}_{m.i.}(u)\ ,
\end{align}
where the vector form factor has been redefined at order $\alpha$ as follows,
\begin{align}\label{eq:dff}
F^{'}_{V}(u)&= F_{V}+\frac{\alpha}{4\pi}f^{m.d.}_{+}(u)\ .
\end{align}
This fact is indicated by a prime on the amplitudes of eq.(\ref{eq:ctl}). As a consequence of eq.(\ref{eq:dff}), it is assumed that $F^{'}_{V}(u)$ is extracted from 
the experiment and hence gives information about the structure dependence and also helps to select the best model or theory that describes 
this form factor and its complications.\\
The same proccedure is applied when the tree level amplitude  depends on two form factors, in this case  the model dependent amplitude can be written as 
follows
\begin{align}\label{eq:2ff}
\mathcal{M}^{v}_{md}&=G_{F}V_{CKM}C_{CG}l_{\mu}\frac{\alpha}{4\pi}[F^{m.d}_{+}(s,u)t^{\mu}_{-}+F^{m.d}_{-}(s,u)t^{\mu}_{+}]\ .
\end{align}
In this general case the form factors $F^{m.d.}_{+}(s,u)$ and $F^{m.d}_{-}(s,u)$ that represent the mixing of hadronic and e.m. effects are  redefined as follows
\begin{align}\label{eq:lchpt}
F^{'}_{+}(s,u)&=F_{+}(s) + \frac{\alpha}{4\pi}F^{m.d.}_{+}(s,u)\ , \nonumber\\
F^{'}_{-}(s,u)&=F_{-}(s) + \frac{\alpha}{4\pi}F^{m.d.}_{-}(s,u)\ .
\end{align}

\section{Model Independent  Radiative Corrections}
In this section the one loop and real photon  corrections to $\tau^{+}\to P^{+} P\nu$ are computed following the techniques explained in \cite{Yena}. A characteristic of this 
correction is  that it is specified completely by the momenta and spin states of the initial and final particles. 
\subsection{Virtual photons.}
The MI one loop electromagnetic corrections are obtained from the diagrams shown in fig.(2) with the tree level amplitude  given in  eq.(\ref{eq:mglevel}).
\begin{figure}[hbt]
\centerline{\includegraphics[width=.95\linewidth]{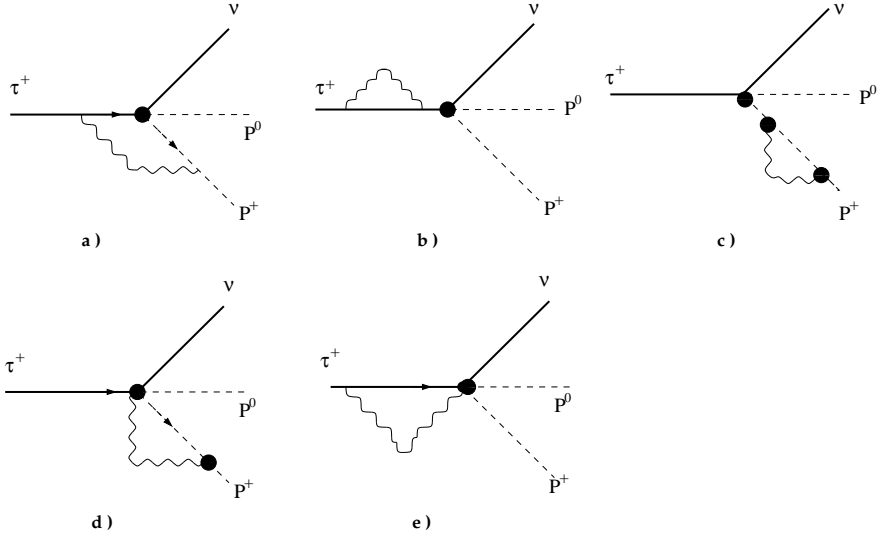}}
\caption{\footnotesize The so-called external diagrams, where a photon line conects external charged legs correspond to a), b) and c), where $P^{+}(P^{0})$ denotes the 
charged(neutral) scalar meson.}
\label{fig2}
\end{figure}
The amplitude of the first external diagram (fig.2a) reads
\begin{align}
\mathcal{M}_{a}&=e^2\bar{u}_{\nu}(q)G\frac{-\imath}{(2\pi)^4}\int dk^4\left[ F_{+}(s')(\sh{p}^{+}-\sh{p}^{0}+\sh{k})\right.\nonumber\\
&\left.+F_{-}(s')(\sh{p}^{+}+\sh{p}^{0}+\sh{k})\right]\gamma^{7}\frac{\sh{p}_{\tau}+\sh{k}-m_{\tau}}{k^2[(p_{\tau}+k)^2-m^2_{\tau}]}\frac{[2\sh{p}^{+}+\sh{k}]}{(p^{+}+k)^2-m^2_{+}} u_{\tau}(p_{\tau})
\end{align}
where now $s'=(p^{0}+p^{+}+k)^2$ and here we define $G=G_{F}V_{CKM}C_{CG}$ in order to avoid long expresions. According to our aim and following the result 
given in \cite{Yena}, the amplitude is separated in one piece that depends only on general QED properties and  a second  piece that depends on the description of the 
structure. This assumption allows us to write the amplitude $\mathcal{M}_{a}$ in the following form 
\begin{align}\label{eq:ma}
\mathcal{M}_{a}&= \mathcal{M}^{(0)}_{\tau}\frac{-\imath e^2}{(2\pi)^4}\int\frac{(2p_{\tau}+k)\cdot (2p^{+}+k)  dk^4 }{k^2[(p_{\tau}+k)^2-m^2_{\tau}][(p^{+}+k)^2-m^2_{+}]}\nonumber\\&
+G\bar{u}_{\nu}(q)h_{\nu}\gamma^{\nu}\gamma^{7}\frac{\imath e^2}{(2\pi)^4}\int \frac{\frac{1}{2}[\gamma^{\alpha},\sh{k}](2p^{+}+k)_{\alpha}}{k^2[(p_{\tau}+k)^2-m^2_{\tau}]}\frac{dk^4}{[(p^{+}+k)^2-m^2_{+}]}u_{\tau}(p_{\tau})\nonumber\\& 
+ \mathcal{M}^{r, md}_{a}
\end{align}
The infrared singularity, some UV divergences  and the major finite correction come from the first line that corresponds to the convection term meanwhile the second line 
corresponds to the spin term which is UV and IR finite \cite{Yena}. The rest of the amplitude $\mathcal{M}^{r, md}_{a}$ can be written as indicated in eq.({\ref{eq:2ff})
and contains all details of strong interactions. From here we assume that the model dependent part is included in the redefinition of the form factors as given in 
eq.(\ref{eq:lchpt}).\\ 
The  convection amplitude, first line in eq.(\ref{eq:ma}), is written\footnote{The integration is done using the FeynCal\cite{hahn} package for Mathematica and the Passarino-Veltman functions can be evaluated with LoopTools. In order to see 
the IR singularity cancellation we use  analytical expression given in the Appendix.} in terms of the tree level amplitude times the electromagnetic correction as follows,
\begin{align}\label{eq:mcc}
\mathcal{M}^{cc}_{a}&= \mathcal{M}^{(0)}_{\tau}\frac{\alpha}{4\pi}\left[ B0[m^2_{\tau},0,m^2_{\tau}] -B0[u,m^2_{+},m^2_{\tau}]+ B0[m^2_{+},0,m^2_{+}]\right.\nonumber\\
&\left.+ 4p_{\tau}\cdot p^{+} C0[m^2_{+},u,m^2_{\tau},\lambda^2] \right]\ .
\end{align}
The convection contribution of the scalar meson self-energy\footnote{See the ref.\cite{Yena}}, fig.(1c), is given by
\begin{align}\label{eq:sep}
\mathcal{M}_{SE}^{cc}&=\mathcal{M}^{(0)}_{\tau}\frac{1}{2}\frac{4\imath\pi\alpha}{(2\pi)^4}\int \frac{dk^4(2p^{+}+k)\cdot(2p^{+}+k)}{k^2[(p^{+}+k)^2-m^2_{+}]^2}\ .
\end{align}
The very well know lepton self-energy, fig.(1b),  reads\footnote{In dimensional regularization $\Delta_{UV}=\frac{2}{4-D}-\gamma_{E}+\ln[4\pi]$.}
\begin{align}\label{eq:sel}
\mathcal{M}_{SE}^{\tau}&=\mathcal{M}^{(0)}_{\tau}\frac{\alpha}{4\pi}\frac{1}{2}\left[-\Delta_{UV}+\ln\left[\frac{m^2_{\tau}}{\mu^2}\right]-4
-2\ln\left[\frac{\lambda^2}{m^2_{\tau}} \right] \right]\ ,
\end{align}
where $\lambda$ is the fictitious mass of the photon  used in order to control the IR divergence and $\mu$ is the well known mass parameter introduced in the 
dimensional regularization method. The total convection amplitude is the sum of (\ref{eq:mcc},\ref{eq:sep},\ref{eq:sel}) and reads
\begin{align}\label{eq:oneloop}
\mathcal{M}_{m.i.}^{cc}&=\mathcal{M}^{(0)}_{\tau}\frac{\alpha}{4\pi}f^{v}_{\rm m.i.}\nonumber\\
&=\mathcal{M}^{(0)}_{\tau}\frac{\alpha}{4\pi} \left[ 2\ln\left[\frac{m_{+}m_{\tau}}{\lambda^{2}}\right]+\frac{m_{+}m_{\tau}}{u}\left[\frac{1}{x_{t}}-x_{t}\right]\ln[x_{t}]
\right.\nonumber\\
&\left.+2m^{2}_{+}y_{t}C0[m^{2}_{+},u,m^{2}_{\tau},\lambda^{2}]-\frac{m^{2}_{+}}{2u}\left(1-r_{t}\right)\ln[r_{t}]-1  \right]
\end{align}
The notation is the same that in \cite{cirigliano} and the expression of 
$C0[m^{2}_{+},u,m^{2}_{\tau},\lambda^{2}]$ is obtained from the general form given in \cite{denner}.
The total convection amplitude is free of UV singularities and is also electromagnetic gauge invariant which  can be  easily checked by adding to the photon propagator 
the term $\xi k^{\mu}k^{\nu}/k^2$, where $\xi$ is an arbitrary parameter and then seeing that $\xi$-dependent contributions from the lepton self energy and convection meson 
self energy cancel the respective terms coming from eq.(\ref{eq:mcc}).\\
The second line of eq.(\ref{eq:ma}) is found to be
\begin{align}\label{eq:bmi}
\mathcal{M}_{2}&=\mathcal{M}^{(0)}_{\tau}\frac{\alpha}{4\pi}(2p_{\tau}\cdot p^{+})f^{r}
+ G\bar{u}_{\nu_{\tau}}(q)h_{\nu}\gamma^{\nu}\gamma^{7}\sh{p}^{+}u(p_{\tau})\frac{\alpha}{4\pi}[2m_{\tau}f^{r}]
\end{align}
This piece is gauge invariant by itself and free of IR and UV singularities and the function $f^{r}$ reads
\begin{align}
 f^{r}&=\frac{1}{m^2_{+}(y_{\tau}^2-4r_{\tau})}\left[ \frac{m^2_{+}(2-y_{\tau})}{u}\left[ \frac{2-y_{\tau}}{2}\ln[r_{\tau}]-\frac{x_{\tau}}{1-x^2_{\tau}}\frac{\ln[x_{\tau}]}{\sqrt{r_{\tau}}}(y_{\tau}^2-4r_{\tau})\right]
\right.\nonumber\\&\left.-\ln[r^2_{\tau}] \right]\ .
\end{align}
In order to get a complete  gauge invariance result, the total MI amplitude  must contain the sum of eqs.(\ref{eq:oneloop},\ref{eq:bmi}), thus the correction to the 
differential tau decay width  is written as follows
\begin{align}\label{eq:dw2f}
\frac{d\Gamma_{P^{+}P^{0}}}{dsdu}&=\frac{|G|^2}{(4\pi)^3m^3_{\tau}}\left[\rho^{(0)}(s,u)\left[1+\frac{\alpha}{2\pi}(f^{v}_{m.i.}+2p_{\tau}\cdot p^{+}f^{r})\right]\right.\nonumber\\
&\left. +\frac{\alpha}{2\pi}\left[m^2_{\tau}f^{r}\right]\left\{ |F_{+}(s)|^2 \mathcal{G}(s,u)+2F^{\dagger}_{+}(s)F_{-}(s)\mathcal{H}(s,u)
\right.\right.\nonumber\\& \left.\left. + |F_{-}(s)|^2 \mathcal{E}(s,u) \right\} \right]
\end{align}
where
\begin{align}
\mathcal{G}(s,u)&=s(m^2_{+}-u)+m^2_{+}(4u-4m^2_{0})+m^2_{\tau}(m^2_{0}-m^2_{+})\ ,\nonumber\\
\mathcal{E}(s,u)&=s(m^2_{+}-u)+m^2_{\tau}m^2_{0}-m^2_{+}m^{2}_{\tau}\ ,\nonumber\\
\mathcal{H}(s,u)&=s(m^2_{+}+u)-m^2_{\tau}m^2_{0}-m^2_{+}m^2_{\tau}\ .
\end{align}
It is worth to mention  that in eq.(\ref{eq:dw2f}),  the function $f^{v}_{m.i.}$ contains all the IR singularity of the one loop correction which is cancelled after taking 
into account the real-photon emission \cite{bloch}.\\
In the case of two pions in the hadronic final state, where the vector form factor is the dominant, the formula (\ref{eq:dw2f}) reads
\begin{align}\label{eq:dw1f}
\frac{d\Gamma_{\pi\pi}}{dsdu}&=\frac{||GF_{V}(s)||^2}{(4\pi)^3m^3_{\tau}}\left[[1+\frac{\alpha}{2\pi}(f^{v}_{m.i.}
+2p_{\tau}\cdot p^{+}f^{r})]\mathcal{D}(s,u)+\frac{\alpha}{2\pi}\big(m^2_{\tau}f^{r}\big)\mathcal{G}(s,u) \right]
\end{align}

\subsection{Real Photon Emission}
In order to be consistent with the work done in the one loop correction, the model independent part of the radiative process must be defined, a task  that is
achieved by using the  low-energy theorems \cite{low}\cite{kroll}.\\ 
According to the Low Theorem \cite{low}, the radiative invariant amplitude denoted as $\mathcal{M}^{\gamma}$ can be expanded  in powers of the photon energy $k$ for 
small $k$ as follows
\begin{align}\label{eq:radtv}
\mathcal{M}^{\gamma}&=\frac{\mathcal{M}}{k} + \mathcal{M}_{1} k^{0}+ k\mathcal{M}_{2}+_{\cdots}
\end{align}
where  the dots symbolize terms with powers of order $\geq 2$ in $k$. The first piece $\mathcal{M}$ (Low-term) and $\mathcal{M}_{1}$  can be calculated completely from the 
non-radiative invariant amplitude meanwhile $\mathcal{M}_{2}$ and the next elements on the serie depend on the theoretical model that describes the details of the 
photon emission from either hadronic external lines or an internal hadronic vertex. This means that eq.(\ref{eq:radtv}) establishes the definition of model independent and model dependent term in the radiative 
amplitude.\\
On the other hand, it was shown  in \cite{kroll} that the unpolarized and squared amplitude
of the radiative process can be splitted into two parts,  one element  of order $1/k^2$ that comes entirely from the Low-term and the rest that contains 
contributions of order $k^{0},k^{1},_{\cdots}$ as it is indicated  in the following equation
\begin{align}
\sum_{spins}|\mathcal{M}^{\gamma}|^2&= \sum_{spins}|\mathcal{M}^{(0)}|^2e^2 \left[\frac{p^{+}\cdot \epsilon(k)}{p^{+}\cdot k}
-\frac{p_{\tau}\cdot\epsilon(k)}{p_{\tau}\cdot k}\right]^2\nonumber\\
& +\sum_{spins}|\mathcal{M}^{'}|^{2}k^{0}+\sum_{spins}|\mathcal{M}^{''}|^{2}k^{1} +_{\cdots} \ ,
\end{align}
where $\epsilon (k)$ is the photon polarization vector and$\sum_{spins}$ indicates an average over initial spin states and a sum over final spin states, except over 
the photon degrees of freedom. The first piece is the Low-term which is precisely the convection term in the MI scheme  and encloses the 
appropriated terms that  cancels  the IR singularity in eq.(\ref{eq:oneloop}). In this work we 
considere only the Low-term for the  radiative amplitude\footnote{The  term of order $k^{0}$ includes contributions from the interference between the model
independent and the model dependent term, therefore it is model dependent.}.\\
According to the previous lines our gauge invariant MI radiative amplitude for the $\tau^{+}\to P^{+}P^{0}\nu \gamma$ decay  reads
\begin{equation}\label{eq:bk}
\mathcal{M}^{\gamma}_{\tau}=\mathcal{M}^{(0)'}_{\tau} e \left[\frac{p^{+}\cdot \epsilon(k)}{p^{+}\cdot k}-\frac{p_{\tau}\cdot\epsilon(k)}{p_{\tau}\cdot k}\right]\ .
\end{equation} 
The soft photon approximation\cite{Denner} was computed in a previous work \cite{francisco} with a careful handling of the infrared singularity \cite{coester} and it 
was shown that the radiative correction depends on  a cutoff energy $\omega_{0}$\footnote{An asumption of the soft photon approximation is that in an experiment, there is a  
minimal energy $\omega_{0}$ for detecting a real photon}. However in this work we considere an alternative procedure \cite{ginsbergI},\cite{ginsbergII},\cite{augusto}
that proposses a separation of the Dalitz-Plot region in such a way that the uncomfortable dependence is avoided.\\
In computing the well known invariant integrals for the real photon correction, a novel approach was done in the work of A. Mart\'inez \textit{et al} \cite{alfonso}, 
obtaining the same result that Ginsberg \cite{ginsbergII}, however  we follow the technique of the later.\\
In the radiative process $\tau\to K\pi\nu\gamma$ a new variable arises, known as the invariant mass of the undetected particles and denoted by $x=(q+k)^2$ with
\begin{align}
x_{-}(s,u)\leq x \leq x_{+}(s,u)
\end{align}
where $k=(k_{0},\vec{k})$ are the energy and the momentum of the photon and the maximal and minimal value of $x$ are given in the Appendix. 
In order to compute the  real photon contribution some assumptions have to be considered, in this respect we adopt those given in \cite{alfonso} which means the following:
\begin{itemize}
\item The allowed kinematical region of all values for $(u,s)$ that satisfy the relation of the three-body
space phase, given by eqs.(\ref{fase1},\ref{fase2}).
\item The values of the Lorentz invariant $x$ consistent with the first point.
\end{itemize}
In other words we considere the set of values of $u,s$ and $x$ that defines a region whose borders are given as follows
\begin{align}\label{eq:ps}
D_{III}&=\left\{ u_{-}(s)\leq u \leq u_{+}(s)\ , s_{min}\leq s \leq s_{max}\ ,\lambda^2\leq x \leq x_{+}(s,u) \right\}
\end{align}
It is important to point out that the IR divergence is precisely within this region. Here the minimal value of $x$ is written in terms of $\lambda$,
the fictitious mass of the photon, the same regulator used in handling the IR singularity in eq.(\ref{eq:oneloop}). As a consecuence of these remarks, it 
is assumed a discrimination of real photons  that could be done in an experimental setup by means of an analysis of the 4-body
radiative Dalitz Plot(See fig.3)\cite{citar tesis de alain}.
\begin{figure}[hbt]
\centerline{\includegraphics[width=.75\linewidth]{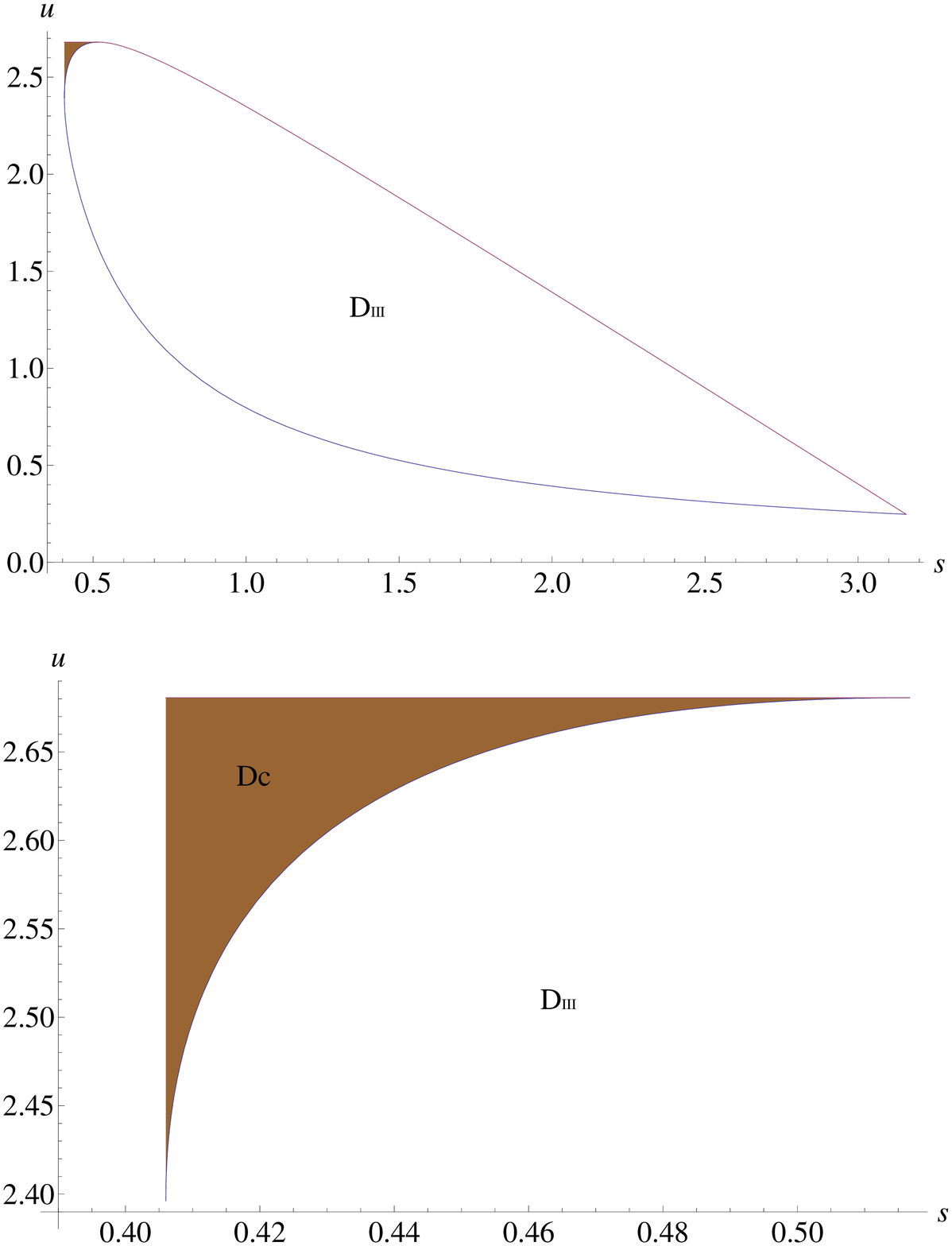}}
\caption{\scriptsize}{\footnotesize The upper plot shows the projection of the 4-body radiative Dalitz Plot onto the u-s plane where can be seen that the 3-body Dalitz
Plot is inside this region. The lower plot shows an amplification of the projected complementary region $D_{c}$ (see text) accesible only to the radiative process.}
\label{fig3}
\end{figure} 
The discriminated photons are inside the complementary region $D_{c}$, only accesible to the radiative process, where the energy of the photon is never zero and as 
a consecuence, the radiative amplitude is free of IR singularity. This region is defined for the following set of values,
\begin{align}
D_{c}&= \left\{ u_{+}\leq u\leq(m_{\tau}-m_{\pi})^2\ ,x_{-}\leq x \leq x_{+}  \ ,\right.\nonumber\\
&\left. s_{min}\leq s\leq \frac{m_{\tau}(m_{\tau}m_{\pi}+m^2_{K}-m^2_{\pi})}{m_{\tau}-m_{\pi}}\right\} 
\end{align}
Once it has been  clarified our assumptions, we can  compute the contribution of the Low term eq.(\ref{eq:bk}) with the corresponding  phase space described in eq.(\ref{eq:ps}).
Notice that we are computing the eq.(18) given in \cite{ginsbergII} but for the $\tau\to K\pi\nu\gamma$ process.\\
The  differential radiative decay width reads 
\begin{align}\label{eq:rfc}
\frac{d\Gamma^{\gamma}_{K\pi}}{dsdu}&=\frac{G^{2}_{F}|V_{us}F^{K\pi}_{+}(0)|^{2}C^{2}_{CG}\rho(s,u)}{m^3_{\tau}(4\pi)^3}\frac{\alpha}{\pi}(I_{1,1}+I_{2,0}+I_{0,2})\ .
\end{align}
The very well know invariant integrals are given in \cite{ginsbergII}, details of the computation and the notation are presented in the appendix, here we just write 
the result in our case as follows,
\begin{align}\label{eq:defI}
I_{2,0}&=\ln \big[\frac{m^2_{\tau}-s}{x_{+}(s,u)}\big]+\ln \big[\frac{\lambda}{m_{\tau}}\big]\ ,\nonumber\\
I_{0,2}&=\ln \big[\frac{m^2_{\tau}-s-u+m^{2}_{K^{0}}}{x_{+}(s,u)}\big]+\ln \big[\frac{\lambda}{m_{\pi^{+}}}\big]\ ,\nonumber\\
I_{1,1}&=\frac{x_{\tau}y_{\tau}}{\sqrt{r_{\tau}}(1-x_{\tau}^{2})}\left\{ Li_{2}\left[ \frac{-a^{2}}{4r_{+}}\right]
-Li_{2}\left[ \frac{-4r_{-}}{a^{2}}\right]  -\ln[x_{\tau}]\ln\left[ \frac{y^{2}_{\tau}-4r_{\tau}}{\sqrt{r_{\tau}}}\right]\right.\nonumber\\& \left.
+\ln[x_{\tau}]\ln\left[ \frac{\lambda^{2}}{m_{\tau}m_{\pi^{+}}}\right] -\ln[x_{\tau}]\ln\left[\frac{x_{\tau}x^{2}_{+}(s,u)}{4r_{+}} \right] \right\}\ . 
\end{align}

\section{Correction to the $\tau^{+}\to K^{0}\pi^{+}\nu$ decay}
The results of the previous sections are applicable to the hadronic final modes ($K^{0}\pi^{+},K^{+}\pi^{0},\pi^{0}\pi^{+}$), however we are interested only in the 
first mode. Consequently we present here the MI electromagnetic corrections to the differential width decay, due to virtual photons eq.(\ref{eq:dw2f}) and real photons 
eq.(\ref{eq:rfc}) computed within the region eq.(\ref{eq:ps}), 
\begin{align}\label{eq:total}
\frac{d\Gamma_{K^{0}\pi^{+}}}{dsdu}&=\frac{G^{2}_{F}|V_{us}F^{K\pi}_{+}(0)|^{2}C^{2}_{CG}}{(4\pi)^{3}m^{3}_{\tau}}\left\{|\tilde{F}_{+}(s)|^{2}\tilde{\mathcal{D}}(s,u)
+2\tilde{F}_{-}(s)\tilde{F}_{+}(s)\tilde{\mathcal{D}}_{3}(s,u)\right.\nonumber\\
&\left.+|\tilde{F}_{-}(s)|^{2}\tilde{\mathcal{D}}_{2}(s,u)\right\}\ ,
\end{align}
where $C^2_{CG}=1/2$ and the $\tilde{\mathcal{D}}$-functions corrected are given by    
\begin{align}\label{eq:tds}
\tilde{\mathcal{D}}(s,u)&=\mathcal{D}(s,u) +\frac{\alpha}{2\pi}\left\{\mathcal{D}(s,u)f^{m.i.}_{I}+\mathcal{G}(s,u)f^{m.i.}_{II}\right\}\nonumber\\
\tilde{\mathcal{D}}_{2}(s,u)&=\mathcal{D}_{2}(s) +\frac{\alpha}{2\pi}\left\{\mathcal{D}_{2}(s)f^{m.i.}_{I}+\mathcal{E}(s,u)f^{m.i.}_{II}\right\}\nonumber\\
\tilde{\mathcal{D}}_{3}(s,u)&=\mathcal{D}_{3}(s,u) +\frac{\alpha}{2\pi}\left\{\mathcal{D}_{3}(s,u)f^{m.i.}_{I}+\mathcal{H}(s,u)f^{m.i.}_{II}\right\}\nonumber\\
\end{align}
where
\begin{align}
f^{m.i.}_{I}&= 2[I_{2,0}+I_{0,2}+I_{1,1}] + f^{v}_{m.i.}+m^2_{\pi^{+}}y_{\tau}f^{r}\ ,\nonumber\\
f^{m.i.}_{II}&= m^2_{\tau}f^{r}\ .
\end{align}
The simple form of eq.(\ref{eq:total}) gives a hand to identify the effect of the e.m. corrections
\begin{itemize}
\item The density function eq.(\ref{eq:density}) is corrected at $\mathcal{O}(\alpha)$ by the M.I. corrections.
\item The form factors includes, for definition in eq.(\ref{eq:lchpt}), the model dependent effects at $\mathcal{O}(\alpha)$.
\end{itemize}
Integrating eq.(\ref{eq:total}) we obtain the total decay width corrected by the model independent e.m. corrections 
\begin{align}
\Gamma_{K^{0}\pi^{+}}&=\Gamma_{K^{0}\pi^{+}}^{(0)}[1+ \delta_{EM}^{m.i.}]
\end{align}
where  the  e.m. correction function reads
\begin{align}\label{eq:fnlc}
\delta_{EM}^{m.i.}&=\frac{[\alpha/2\pi]}{m^{8}_{\tau}I^{(0)}_{K\pi}}\frac{3}{4}\int\left\{|\tilde{F}_{+}(s)|^{2}[\mathcal{D}(s,u)f^{m.i.}_{I}+\mathcal{G}(s,u)f^{m.i.}_{II}]\right.\nonumber\\
&\left. +|\tilde{F}_{-}(s)|^{2}[\mathcal{D}_{2}(s)f^{m.i.}_{I}+\mathcal{E}(s,u)f^{m.i.}_{II}]\right.\nonumber\\
&\left. +2\tilde{F}_{-}(s)\tilde{F}_{+}(s)[\mathcal{D}_{3}(s,u)f^{m.i.}_{I}+\mathcal{H}(s,u)f^{m.i.}_{II}]\right\} dsdu
\end{align}
In order to estimate the electromagnetic correction we follow the theoretical description for the form factors as given in section(2.1), hence we obtain 
$\delta_{EM}^{m.i.}= -0.127\%$. The aim of this work has been achieved by writing eq.(\ref{eq:fnlc}), however the Dalitz plot corrections can be obtained straightforward by
using  eq.(\ref{eq:tds}).\\ On the other hand, Antonelli \textit{et al}\cite{lastC} have given an estimate of the long distance electromagnetic corrections to $\tau^{+}\to K^{0}\pi^{+}\nu$,
where the structure dependent effects have been neglected and their approach relies in the analysis already done in $Kl_{3}$ and $\tau\to \pi\pi\nu$. They found 
$\delta_{em}^{\bar{K}^{0}\pi^{+}}=(-0.15\pm 0.2)\%$ where $0.2$ is the uncertainty they assigned to the unknown structure dependent effects.   

\section{Conclusions}
Summarizing the work done, we  provide the MI electromagnetic corrections to the $\tau^{+}\to K^{0}\pi^{+}\nu$ decay following the procedure described in \cite{Yena}
considering both virtual and real photon within the 3-body phase space region eq.(\ref{eq:ps}). As it has been pointed out this correction is electromagnetic gauge 
invariant, free of IR singularities, free of UV singularities and most important it does not have an UV cutoff. This approach considers that all structure dependence 
is included inside the form factors $\tilde{F}_{\pm}(s)$ after an appropiated redefinition. In this respect our work is not focused in the theoretical description of
the form factors and the corresponding solution to the remaining UV problem.\\ 
On the other hand, it is important to accentuate that eq.(\ref{eq:fnlc}) can be used for the three hadronic final states 
($\pi^{\pm}\pi^{0},\pi^{\pm} K^{0},\pi^{0} K^{\pm}$) with the corresponding changes. Concerning the $\tau$ decay into the 2$\pi$ final state, the universal 
electromagnetic correction given in \cite{cirigliano} is slighty different from the MI correction, we stressed this to be more qualitative than quantitative.\\

\section*{Acknowledgements}
This work was supported in part by 
CONACYT SNI (M\'exico). F.V. Flores Ba\'ez  thank AECID (MAEC, Spain) for grant M\'ex:/0288/09 and also thank the IFIC-Valencia
for their hospitality.

\section*{Appendix.}

\subsection*{Kinematics.}
In the radiative process a new invariant Mandelstam variable is introduced,  $x=(q+k)^2$ whose maximum and  minimum  value are given 
as follows
\begin{align}
x_{\pm}(s,u)&= \frac{1}{2m^2_{\pi}}\left[u(m^2_{+}-m^{2}_{0})-m^2_{+}(m^2_{+}-m^2_{0})+m^2_{\tau}(m^2_{+}+m^{2}_{0}) \right.\nonumber\\&
\left.-s(m^2_{\tau}-m^2_{+}-u)\pm \lambda^{1/2}(u,m^2_{+},m^2_{\tau})\lambda^{1/2}(s,m^2_{+},m^2_{0})  \right]
\end{align}
where the well known  K\"{a}llen function  reads
\begin{align}
\lambda(a,b,c)&= a^2+b^2+c^2 -2(ab+ac+bc)\ .
\end{align}
\subsection*{Invariant amplitudes.}
According to the definition given in \cite{ginsbergII}, the radiative function of the Low term is written as follows 
\begin{align}
g^{\gamma}&= I_{11}+I_{20}+I_{02}
\end{align}
These type of integrals have been already  studied in \cite{ginsbergII}, here we present brief details about computing them for our situation. We start with the easiest one 
which we write it as follows 
\begin{align}\label{eq:I20int}
I_{2,0}&=-\lim_{\lambda\to 0}\int_{\lambda^{2}}^{x_{+}(t,u)}p_{\tau}^{2} i_{2,0}dx\ ,\nonumber\\
i_{2,0}&=\frac{1}{2\pi}\int \frac{d^{3}\vec{k}d^{3}\vec{q}}{k_{0}q_{0}}\frac{\delta(\vec{P}-\vec{k}-\vec{q})\delta(E_{\nu}-k_{0}-q_{0})}{[(p_{\tau}-k)^2-m^2_{\tau}]^{2}}\ .
\end{align}
where $P=p_{\tau}-p_{+}-p_{0}$ and $E_{\nu}=E_{\tau}-E_{+}-E_{0}$. The regulator for the IR singularity is chosen to be the same than the one used in the one loop
corrections which means that $k^2=\lambda^2$ for a real photon with a small fictitious mass. A frame is chosen in such a way that $\vec{P}=0$, hence 
\begin{align}\label{eq:i20}
i_{2,0}&=\frac{1}{2\pi}\int \frac{d^{3}\vec{k}d^{3}\vec{q}}{k_{0}q_{0}}\frac{\delta(-\vec{k}-\vec{q})\delta(E_{\nu}-k_{0}-q_{0})}{4[p_{\tau}\cdot k-\frac{\lambda^2}{2}]^{2}}\ .
\end{align}
Integration over the photon variable  allow us to eliminate one delta function, then eq.(\ref{eq:i20}) reads
\begin{align}
i_{2,0}&=\frac{1}{8\pi}\int \frac{d^{3}\vec{q}/q_{0}}{\sqrt{q^2_{0}+\lambda^{2}}}\frac{\delta(E_{\nu}-\sqrt{q^{2}_{0}+\lambda^{2}}-q_{0})}{\big[E_{\tau}\sqrt{q^{2}_{0}
+\lambda^{2}}+\vec{p_{\tau}}\cdot \vec{q}-\lambda^{2}/2\big]^{2}}\ .
\end{align}
The next integration is computed in spherical coordinates,  first over the azimuth angle $\phi$ and then over $q_{0}$, then last equation reads
\begin{align}
i_{2,0}&=\int_{0}^{\pi} \frac{(E_{\nu}^{2}-\lambda^{2})\sin\theta d\theta}{2[E_{\tau}(E^{2}_{\nu}+\lambda^{2})+|\vec{p}_{\tau}|(E^{2}_{\nu}-\lambda^{2})\cos\theta
-\lambda^{2}E_{\nu}]^{2}}\nonumber\\
\end{align}
Throwing away contributions of equal or greater order than $\mathcal{O}(\lambda^{4})$, the previous expression is found to be
\begin{equation}\label{eq:idos}
i_{2,0}=\frac{(x-\lambda^{2})}{x^{2}m^{2}_{\tau}+4\lambda^{2}(P\cdot p_{\tau})^{2}-2\lambda^{2}p_{\tau}^{2}x}\ .
\end{equation}
Finally the $x$-integration in eq.(\ref{eq:I20int}) is done straightforward
\begin{align}\label{eq:idoskpi}
I_{2,0}&=-\lim_{\lambda\to 0}\int_{\lambda^{2}}^{x_{+}(t,u)}p_{\tau}^{2} i_{2,0}dx=\ln\left[\frac{m^{2}_{\tau}-s}{x_{+}(s,u)}\right]
+\ln\left[\frac{\lambda}{m_{\tau}}\right]\ .
\end{align}
The computation of $I_{0,2}$ is quite similar, here we just present the result
\begin{align}
I_{0,2}&=-\lim_{\lambda\to 0}\int_{\lambda^{2}}^{x_{+}(s,u)}p_{+}^{2} i_{0,2}dx=\ln\left[\frac{m^{2}_{\tau}-u-s+m^{2}_{0}}{x_{+}(s,u)}\right]+\ln\left[\frac{\lambda}{m_{+}}\right]\ .
\end{align}
The third integral  is computed following the same procedure, first we write it as follows
\begin{align}\label{eq:I11}
I_{1,1}&=\lim_{\lambda\to 0}\int_{\lambda^2}^{x_{+}(s,u)} 2p_{\tau}\cdot p_{+}i_{1,1}dx\ ,\nonumber\\
i_{1,1}&=\frac{1}{2\pi}\int \frac{d^{3}\vec{q}d^{3}\vec{k}\delta(\vec{k}+\vec{q})\delta(E_{\nu}-k_{0}-q_{0})}{4q_{0}k_{0}[E_{\tau}k_{0}-
\vec{p_{\tau}}\cdot \vec{k}-\lambda^{2}/2][E_{+}k_{0}-\vec{p_{+}}\cdot \vec{k}+\lambda^{2}/2]}\ .
\end{align}
After integrating the delta function $\delta(\vec{k}+\vec{q})$ we get
\begin{align}\label{eq:i11delta}
i_{1,1}&=\frac{1}{2\pi}\int \frac{d^{3}\vec{q}\delta(E_{\nu}-k_ {0}-q_{0})}{4q_{0}k_ {0}[E_{\tau}k_{0}+\vec{p_{\tau}}\cdot \vec{q}
-\lambda^{2}/2][E_{+}k_{0}+\vec{p_{+}}\cdot \vec{q}+\lambda^{2}/2]}\ ,
\end{align}
where $k_{0}=\sqrt{q^{2}_{0}+\lambda^{2}}$.  The Feynman trick  allows us to combinate  propagators as follows
\begin{align}\label{eq:feynt}
&\frac{1}{[E_{\tau}k_{0}+\vec{p_{\tau}}\cdot \vec{q}-\lambda^{2}/2][E_{+}k_{0}+\vec{p_{+}}\cdot \vec{q}+\lambda^{2}/2]}=\int_{0}^{1}\frac{dz}{[c+\vec{q}\cdot \vec{p}]^{2}}
\end{align}
where now 
\begin{align}
c&=E_{+}k_{0}+z((E_{\tau}-E_{+})k_{0}-\lambda^{2})+\lambda^{2}/2\ ,\nonumber\\
\vec{p}&= \vec{p_{\tau}}z+\vec{p_{+}}(1-z)\ .
\end{align}
After using eq.(\ref{eq:feynt}) and the properties of  the delta function, the Eq.(\ref{eq:i11delta}) reads
\begin{align}\label{eq:i11}
i_{1,1}&=\frac{1}{2\pi}\frac{E_{\nu}^{2}-\lambda^{2}}{2E^{2}_{\nu}}\int_{0}^{1}\int_{0}^{2\pi}\int_{0}^{\pi} \frac{\sin \theta d\theta d\phi dz}{4[\bar{c}+\frac{E_{\nu}^{2}-\lambda^{2}}{2E_{\nu}} |\vec{p}| \cos \theta]^{2}}\nonumber\\
&=\frac{E_{\nu}^{2}-\lambda^{2}}{4E^{2}_{\nu}}\int_{0}^{1}\frac{dz}{\bar{c}^{2}-(|\vec{p}|\frac{E_{\nu}^{2}-\lambda^{2}}{2E_{\nu}})^{2}}\ ,
\end{align}
where 
\begin{align}
\bar{c}&=E_{+}\frac{E_{\nu}^{2}+\lambda^{2}}{2E_{\nu}}+z((E_{\tau}-E_{+})\frac{E_{\nu}^{2}+\lambda^{2}}{2E_{\nu}}-\lambda^{2})+\lambda^{2}/2\ .
\end{align}
In order to do the z-integration, the denominator is written as a polynomial function on z,
\begin{align}\label{eq:coeff}
\bar{c}^{2}-(|\vec{p}|q_{0})^{2}=&\frac{1}{4E_{\nu}^{2}}\left\{ z^{2} \left[ ((E_{\tau}-E_{+})(E_{\nu}^{2}+\lambda^{2})-2E_{\nu}\lambda^{2})^{2}
-\vec{p_{\tau}}^{2}(E^{2}_{\nu}-\lambda^{2})^{2}\right.\right.\nonumber\\
&\left.\left.+2\vec{p_{\tau}}\cdot\vec{p_{+}}(E_{\nu}^{2}-\lambda^{2})^{2}-\vec{p_{+}}^{2}(E_{\nu}^{2}-\lambda^{2})^{2}\right]\right. \nonumber\\
&\left. +2z\left[ E_{+}(E_{\nu}^{2}+\lambda^{2})\left[(E_{\tau}-E_{+})(E_{\nu}^{2}+\lambda^{2})-2\lambda^{2}E_{\nu}\right] \right.\right.\nonumber\\
&\left.\left. +(-\vec{p_{\tau}}\cdot \vec{p_{+}}+\vec{p_{+}}^{2})(E_{\nu}^{2}-\lambda^{2})^{2} \right.\right.\nonumber\\
&\left.\left.  +\lambda^{2}E_{\nu}(E_{\tau}-E_{+})(E_{\nu}^{2}+\lambda^{2}) -2\lambda^{4}E^{2}_{\nu} \right] \right.\nonumber\\
&\left. +\left[ E_{+}^{2}(E_{\nu}^{2}+\lambda^{2})^{2}+\lambda^{4}E_{\nu}^{2}\right.\right.\nonumber\\
&\left.\left. +2\lambda^{2}p_{+}\cdot P(E_{\nu}^{2}+\lambda^{2})-\vec{p_{+}}^{2}(E_{\nu}^{2}-\lambda^{2})^{2}   \right]  \right\}\nonumber\\
&=\frac{1}{4E_{\nu}^{2}}[\alpha z^{2}+2\beta z+\gamma]\ ,
\end{align}
The coeficients, in terms of Lorentz invariant products, are written as follows
\begin{align}\label{eq:cfin}
\alpha=& (x-\lambda^{2})^{2}\left( m^{2}_{\tau}+m^{2}_{+}-2(p_{\tau}\cdot p_{+})   \right)+4\lambda^{4}x\nonumber\\& 
 +4\lambda^{2}(p_{\tau}\cdot P-p_{+}\cdot P)\left(  (p_{\tau}\cdot P-p_{+}\cdot P)-(x+\lambda^{2})\right)\nonumber\\
\beta=&(x-\lambda^{2})^{2}(p_{\tau}\cdot p_{+}-m^{2}_{+})+2\lambda^{4}\left\{P\cdot p_{\tau}-3P\cdot p_{+}\right\}\nonumber\\
& + 4\lambda^{2}\left\{P\cdot p_{\tau}P\cdot p_{+}-(P\cdot p_{+})^{2}\right\} -2\lambda^{4}x \nonumber\\
&+(x-\lambda^{2})\lambda^{2}\left\{P\cdot p_{\tau}-3P\cdot p_{+}\right\}\nonumber\\
\gamma=&(x-\lambda^{2})^{2}m^{2}_{+}+\lambda^{2}(2P\cdot p_{+})\left(2P\cdot p_{+}+x+\lambda^{2}\right)+\lambda^{4}x
\end{align}
It is straigthforward to find that eq.(\ref{eq:i11}) reads
\begin{align}\label{eq:final}
i_{1,1}&=\int_{0}^{1}\frac{(x-\lambda^{2})dz}{\alpha z^{2}+2\beta z+\gamma}=\frac{(x-\lambda^{2})}{2\sqrt{\beta^{2}-\alpha\gamma}}
\ln\left[ \frac{\beta+\gamma+\sqrt{\beta^{2}-\alpha\gamma}}{\beta+\gamma-\sqrt{\beta^{2}-\alpha\gamma}}  \right]\ .
\end{align}
The coefficients given  in eq.(\ref{eq:cfin}), can be written in terms of the invariant variables $(s,u)$ with the help of the following scalar products
\begin{align}\label{eq:scap}
P\cdot p_{\tau}&=\frac{1}{2}(x+m^2_{\tau}-s)\nonumber\\
P\cdot p_{+}&=\frac{1}{2}(m^2_{\tau}+m^2_{0}-s-u)\nonumber\\
p_{\tau}\cdot p_{+}&=\frac{1}{2}(m^2_{\tau}+m^2_{+}-u)
\end{align}
Then with eq.(\ref{eq:final}) and eq.(\ref{eq:scap}) we write eq.(\ref{eq:I11}) in the conventional form given by Ginsberg\footnote{Eq.(25) in \cite{ginsbergII} },
\begin{align}\label{eq:iuno}
I_{1,1}&=\lim_{\lambda\to 0}\int_{\lambda^{2}}^{x_{+}}dx\frac{E_{\tau}}{\bar{\delta}}\ln\left[\frac{x^{2}E_{\tau}+2\lambda^{2}\epsilon_{\nu}(s-m^{2}_{\tau})
+(x-\lambda^{2})\bar{\delta}}{x^{2}E_{\tau}+2\lambda^{2}\epsilon_{\nu}(s-m^{2}_{\tau})-(x-\lambda^{2})\bar{\delta}}\right]\nonumber\\
\end{align}
where
\begin{align}\label{eq:eqs}
\bar{\delta}&=\sqrt{x^{2}|\vec{p}_{\tau}|^{2}+\lambda^{2}a^{2}}\ , \qquad E_{\tau}=\frac{m^2_{\tau}+m^2_{+}-u}{2m_{+}}\nonumber\\
a^2&= -x_{+}(s,u)x_{-}(s,u)\ , \qquad |\vec{p_{\tau}}|=\frac{\sqrt{\lambda(u,m^2_{\tau},m^2_{+})}}{2m_{+}}\nonumber\\ 
\epsilon_{\nu}&=\frac{u+s-m^2_{\tau}-m^2_{0}}{2m_{+}}
\end{align}
Recalling the definition  given by Ginsber for $r_{\pm}$, 
\begin{align}
r_{\pm}&=\frac{x_{\tau}}{m_{\tau}}\left\{\Omega_{+}\pm \sqrt{ \Omega_{+}^{2} -\frac{1}{16}m^{2}_{\tau}a^{4}} \right\}\nonumber\\
\Omega_{+}&=\epsilon_{\nu}|p_{\tau}|^{2}(s-m^{2}_{\tau})-\frac{1}{4}a^{2}E_{\tau}\ .
\end{align}
and using the master formula\footnote{Eq.(28) in\cite{ginsbergII}},  we get the final result
\begin{align}
I_{1,1}&=\frac{x_{\tau}y_{\tau}}{\sqrt{r_{\tau}}(1-x_{\tau}^{2})}\left\{ Li_{2}\left[ \frac{-a^{2}}{4r_{+}}\right]
-Li_{2}\left[ \frac{-4r_{-}}{a^{2}}\right] +\ln[x_{\tau}]\ln\left[ \frac{\lambda^{2}}{m_{\tau}m_{+}}\right]  \right.\nonumber\\& \left.
-\ln[x_{\tau}]\ln\left[ \frac{y^{2}_{\tau}-4r_{\tau}}{\sqrt{r_{\tau}}}\right]-\ln[x_{\tau}]\ln\left[\frac{x_{\tau}x^{2}_{+}(s,u)}{4r_{+}} \right] \right\}
\end{align}
where
\begin{align}
r_{\tau}&=\frac{m^{2}_{\tau}}{m^{2}_{+}}\ ,\qquad y_{\tau}=1+r_{\tau}-\frac{u}{m^{2}_{+}}\nonumber\\
x_{\tau}&=\frac{1}{2\sqrt{r_{\tau}}}\left[y_{\tau}-\sqrt{y^{2}_{\tau}-4r_{\tau}} \right]
\end{align}
Notice this result is useful also for the $\tau^{+}\to K^{+}\pi^{0}\nu\gamma$ decay, where we must put $m_{+}=m_{K^{+}}$ and $m_{0}=m_{\pi^{0}}$.\\
\subsection*{Scalar function C0.}
In the one loop calculation, it is found the 3-point scalar function C0 that can be evaluated by  LoopTools. However, in order to see
the cancellation of the IR singularity, the C0 function is written in terms of logarithms and dilogarithms with the help of the general form \cite{denner}, 
\begin{align}
C0[m^{2}_{+},u,m^{2}_{\tau},\lambda^{2}]&=C0[m^{2}_{+},u,m^{2}_{\tau},\lambda^{2},m^{2}_{+},m^{2}_{\tau}]\nonumber\\
&=\frac{x_{t}}{m_{+}m_{\tau}(1-x^{2}_{t})}\left[ -\frac{1}{2}\ln^{2}[x_{t}]-\frac{\pi^{2}}{6}+2\ln[x_{t}]\ln[1-x^{2}_{t}]\right.\nonumber\\
&\left.+\frac{1}{8}\ln^2[r_{t}]  + Li_{2}[x^{2}_{t}]+ Li_{2}\left[1-\frac{x_{t}}{\sqrt{r_{t}}}\right]+ Li_{2}[1-x_{t}\sqrt{r_{t}}]\right.\nonumber\\
& \left.-\ln[x_{t}]\ln\left[\frac{\lambda^{2}}{m_{+}m_{\tau}}\right]\right]
\end{align}
The $B0$ function is very well known and all the specific cases can be obtained easily from its definition\cite{bo}. 

\input{bib_samplefvflores}

\end{document}